\documentstyle[aps,psfig,epsf,multicol]{revtex}
\draft
\begin{document}
\title{Laser induced breakdown of the magnetic field reversal symmetry in the propagation of unpolarized light}
\author{G. S. Agarwal and Shubhrangshu Dasgupta}
\address{Physical Research Laboratory, Navrangpura, Ahmedabad-380 009, India}
\date{\today}
\maketitle

\begin{abstract}
We show how a medium, under the influence of a coherent control field which is resonant or close to resonance to an appropriate atomic transition, can lead to very strong asymmetries in the propagation of unpolarized light when the direction of the magnetic field is reversed. We show how EIT can be used to mimic 
 effects occurring in natural systems and that EIT can produce very
large asymmetries as we use electric dipole allowed transitions. Using
density matrix calculations we present results for the breakdown of the magnetic field reversal symmetry for two different atomic configurations.
\end{abstract}
\begin{multicols}{2}
\section{Introduction}
It is well known how an isotropic medium becomes anisotropic by the application
of a magnetic field \cite{landau}. In the special case when the magnetic field is parallel to
the direction of the applied field and if we include the electric dipole 
contribution to susceptibilities, then the two circularly polarized light waves
travel independently of each other. The propagation itself is determined by the 
magnetic field dependent optical susceptibilities $\chi_\pm$. If we write the
incident field of frequency $\omega$ in the form
\begin{equation}
\vec{E}\equiv ({\cal E}_+\hat{\epsilon}_++{\cal E}_-\hat{\epsilon}_-)e^{ikz-i\omega t}+\mathrm{c.c.},
\label{efield}
\end{equation}
where
\begin{equation}
\hat{\epsilon}_\pm=\left(\frac{\hat{x}\pm i\hat{y}}{\sqrt{2}}\right),~~{\cal E}_\pm=\left(\frac{{\cal E}_x\mp i{\cal E}_y}{\sqrt{2}}\right),~~ k=\frac{\omega}{c}, 
\label{pols}
\end{equation}
the output field is given by 
\begin{equation}
\vec{E}_0=\vec{\cal E}_0e^{ikz-i\omega t}+\mathrm{c.c.},
\label{outfield}
\end{equation}
where
\begin{equation}
\vec{\cal E}_0={\cal E}_+\hat{\epsilon}_+e^{2\pi ikl\chi_+}+{\cal E}_-\hat{\epsilon}_-e^{2\pi ikl\chi_-}.
\label{outamp}
\end{equation}
The susceptibilities $\chi_\pm$ also depend on the frequency of the applied 
field. The rotation of the plane of polarization \cite{mor,mor2,gaeta,stenholm,mor1,berkley} and the dichroism can be 
calculated in terms of the real and imaginary parts of $\chi_\pm$. An interesting
situation arises if the incident pulse is unpolarized. In that case there is a 
random phase difference between ${\cal E}_x$ and ${\cal E}_y$ and the intensities
along two orthogonal directions are equal, i.e.,
\begin{equation}
\langle {\cal E}_x^*{\cal E}_x\rangle = \langle{\cal E}_y^*{\cal E}_y\rangle =\frac{I}{2},~~~~\langle{\cal E}_x^*{\cal E}_y\rangle=0,
\label{unpol1}
\end{equation}
$I$ being the intensity of the incident pulse.
From Eqs. (\ref{pols})-(\ref{unpol1}), we can evaluate the output intensity 
$I_0=\langle |\vec{\cal E}_0|^2\rangle$:
\begin{eqnarray}
I_0&\equiv &\frac{I}{2}\left[|e^{2\pi ikl\chi_+}|^2+|e^{2\pi ikl\chi_-}|^2\right]\nonumber\\
&=&\frac{I}{2}\left[\exp{\{-4\pi kl\mathrm{Im}(\chi_+)\}}+\exp{\{-4\pi kl\mathrm{Im}(\chi_-)\}}\right].
\label{outint1}
\end{eqnarray}
Thus the output intensity is a symmetric function of $\chi_+$ and $\chi_-$. If the
susceptibilities $\chi_\pm$ obey the following relation when the direction of 
the magnetic field is reversed
\begin{equation}
\chi_\pm (B)=\chi_\mp (-B),
\label{symmchi}
\end{equation}
then
\begin{equation}
I_0(B)=I_0(-B).
\label{equal1}
\end{equation}
Thus for unpolarized light the output intensity is the same whether the magnetic 
field is parallel or antiparallel to the direction of propagation of the 
electromagnetic field as long as (\ref{symmchi}) is satisfied. 

In this paper we investigate if the transmission of unpolarised light
through an otherwise isotropic medium can be sensitive to the direction of
the magnetic field. We demonstrate how a suitably applied control field
could make the transmission dependent on the direction of the magnetic
field. This is perhaps the first demonstration of the
dependence of transmission of the unpolarised light on the direction of
$B$. For a large range of parameters we find that transmission could be
changed by a factor of order two. We also report a parameter domain where
the medium becomes opaque for one direction but becomes transparent for
the reversed direction of the magnetic field. This work is motivated by the
phenomena of optical activity and the magneto-chiral anisotropy which
occur in many systems in nature \cite{raupach,nature,rupa}. The latter effect has recently
become
quite important. Several ingenious measurements of this effect have been
made though
the effect in natural systems is quite small. The smallness of the effect
arises from the fact that the effect involves a combination of electric
dipole, magnetic dipole and quadrupole effects[11-14]. The magnetochiral
anisotropy
is just the statement $I(B)$ not equal to $I(-B)$. The effect we report is
analogous but quite different in its physical content.

In this paper we show how rather large asymmetry between $I_0(B)$ and $I_0(-B)$
can be produced by using a coherent control field. The asymmetry could be large as we use only the electric dipole transitions. To demonstrate the idea we 
consider different specific situations depending on the transition on which
the control field is applied. The control 
field can be used to modify, say, $\chi_+$ leaving $\chi_-$ unchanged \cite{mor,gaeta,mor1,eit}. Thus in 
presence of the control field we violate the equality (\ref{symmchi}) and this 
can result in large magneto-asymmetry in the propagation of unpolarized light.
 Such a large asymmetry, which we would refer to as magnetic field reversal asymmetry (BRAS in short), is induced by selectively applying the control
field so as to break the time-reversal symmetry. It is important to note that
we work with electric dipole transitions only. The control field is used to 
mimic the effects which occur in nature due to a combination of higher order 
multipole transitions.

The structure of our paper is as follows. In Sec. II, we will discuss how one
can use a control field to create large BRAS. We present a very 
simple physical model.
 We present relevant analytical results. 
In Sec. III, 
we introduce another model, where the control field is applied such that we
get a ladder system. We present the analytical results for BRAS
in such a system. We also discuss the effect of atomic motion on BRAS in Sec. IV.

\section{Large magnetic field reversal asymmetry using EIT}
The application of a coherent pump leads to the well-studied electromagnetically induced transparency
(EIT) \cite{eit} and coherent population trapping (CPT) \cite{cpt}. The usage of
pump and probe in a Lambda configuration is especially useful in suppressing 
the absorption of the probe, particularly if the lower levels of the $\Lambda$
configuration are metastable and if the pump is applied between initially 
unoccupied levels. We explore how EIT can help in producing large BRAS.
We first explain the basic idea in qualitative terms and then would produce 
detailed results using density matrix equations for several systems of interest.

Consider the following scenario. Let us consider first the case when $\vec{B}$
is applied parallel to the direction of propagation of the electromagnetic field.
Suppose the control field is applied such that the $\sigma_+$ component becomes
transparent, i.e., Im$[\chi_+(B)]\approx 0$. For magnetic field bigger than the
typical linewidth,the component $\sigma_-$ is off-resonant. Thus $\sigma_-$ 
exhibits very little absorption Im$(\chi_-)\approx 0$. Under such 
conditions, Eq.\ (\ref{outint1}) shows that the transmitted intensity $\approx I$. Now if the direction of the 
magnetic field is reversed, then we easily find the situation when $\sigma_+$ 
component becomes off resonant
from the corresponding transition, i.e., Im$(\chi_+)\approx 0$; the $\sigma_-$
component can become resonant and suffers large absorption, i.e., exhibits large Im$(\chi_-)$. This gives 
rise to an intensity $\sim I/2$. Thus the transmittivity reduces by a factor $1/2$
upon reversal of the magnetic field. It is thus clear how coherent fields can be
used to create large BRAS.
{\narrowtext
\begin{figure}
\epsfxsize 8cm
\centerline{
\epsfbox{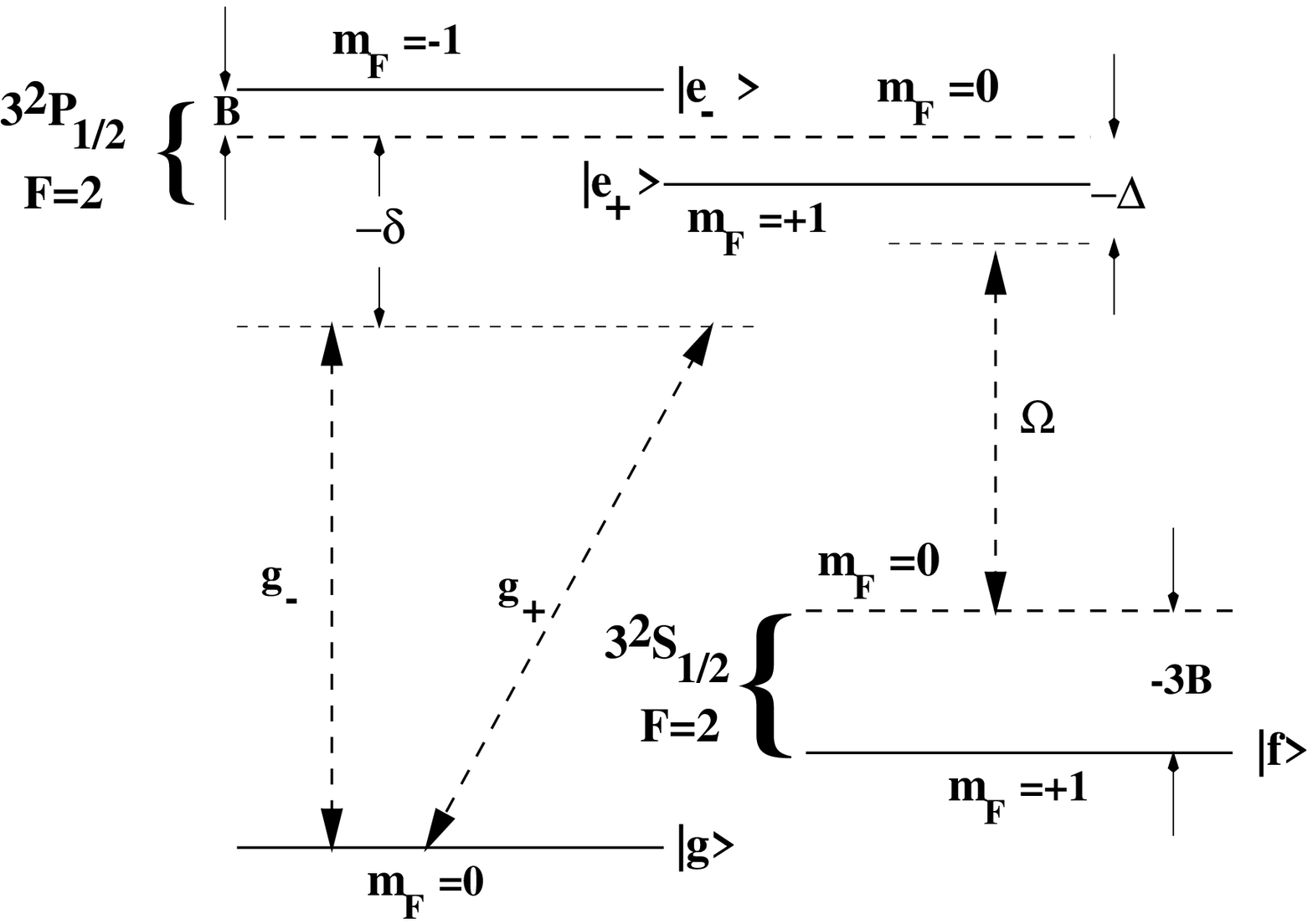}}
\caption{The $^{23}$Na hyperfine level configuration is shown here. Here, $B > 0$ 
is the applied magnetic field strength, $2g_\pm$ are the probe Rabi 
frequencies for the $\sigma_\pm$ components, and $\Omega$ is the half of the 
pump Rabi frequency. The respective detunings $\delta = [\omega_p-\omega_{e_+g}(B=0]$ and $\Delta$ for the 
probe and pump fields are defined with respect to the energy separation between 
the levels $(|3^2P_{1/2}; F=2,m_F=0\rangle, |g\rangle)$ and $(|3^2P_{1/2}; F=2,m_F=0\rangle, |3^2S_{1/2}; F=2,m_F=0\rangle)$, respectively. Changing the direction of the magnetic field interchanges the positions of $|e_-\rangle$ and $|e_+\rangle$. Besides the level $|f\rangle$ moves above the dashed line for $|3^2S_{1/2}; F=2,m_F=0\rangle$.}
\label{config1}
\end{figure}
}
We now demonstrate the feasibility of these ideas. We consider a
configuration [see Fig.\ \ref{config1}] which can be found, for example, in hyperfine levels of $^{23}$Na \cite{bassani}. The level
$|g\rangle$ $(|3^2S_{1/2}; F=1,m_F=0\rangle)$ is coupled to the upper level 
$|e_-\rangle$ $(|3^2P_{1/2}; F=2,m_F=-1\rangle)$ and $|e_+\rangle$ $(|3^2P_{1/2};
 F=2,m_F=+1\rangle)$ by the $\sigma_-$ and 
$\sigma_+$ components of the probe field. The susceptibilities for
the two components of the probe acting on the transitions $|g\rangle\leftrightarrow |e_-\rangle$ and
$|g\rangle\leftrightarrow |e_+\rangle$ are given by
\begin{mathletters}
\begin{eqnarray}
\chi_-(B)&=&\frac{-i\gamma\alpha_0}{i(\delta -B)-\Gamma_{e_-g}},\\
\chi_+(B)&=&\frac{-i\gamma\alpha_0}{i(\delta +B)-\Gamma_{e_+g}}, \delta\equiv \omega - \omega_{e_+g}(B=0),
\end{eqnarray}
\label{chi1}
\end{mathletters}
where, $\alpha_0$ is given by $N|\vec{d}|^2/\hbar\gamma$ and is related to the absorption in the line-center for $B=0$. It should be borne in mind that $B$ 
represents the Zeeman splitting of the level $m_F=-1$. Thus $B$ has the unit of
frequency.
 Here $2\gamma$ is the spontaneous decay rate from the level
$|e_-\rangle$, $\Gamma_{e_-g} =\gamma (\Gamma_{e_+g}=4\gamma/3)$ is the decay rate of 
the off-diagonal density matrix elements between level $|e_-\rangle$ $(|e_+\rangle)$ and 
$|g\rangle$, $N$ is the atomic number density, $|\vec{d}|$ is the dipole
moment matrix element between the levels $|e_-\rangle$ and $|g\rangle$, and $\delta$ is the detuning of the probe field from the $|g\rangle \leftrightarrow |3^2P_{1/2}; F=2,m_F=0\rangle$ 
transition. Note that $\delta$ would always be defined with respect to 
the levels in the absence of the magnetic field. Using the Eqs.\ (\ref{outint1}) and (\ref{chi1}), one easily finds that the relation (\ref{equal1}) holds for all $\delta$. In all equations, $B$ would be considered as positive quantity.

To create a large asymmetry between the output intensities $I_0(B)$ and $I_0(-B)$, 
we now apply a coherent 
control field
\begin{equation}
\vec{E}_p(z,t)=\vec{\cal E}_p(z)e^{-i\omega_p t}+\textrm{c.c.}
\label{pump1}
\end{equation}
on the transition $|e_+\rangle\leftrightarrow |f\rangle$ $(|3^2S_{1/2}; F=2,m_F=+1\rangle)$. 
This modifies the susceptibility $\chi_+$ of the $\sigma_+$ component to 
\begin{eqnarray}
\bar{\chi}_+(B)&=&\frac{-i\gamma\alpha_0[i(\delta-\Delta+3B)-\Gamma_{fg}]}{[i(\delta+B)-\Gamma_{e_+g}][i(\delta-\Delta+3B)-\Gamma_{fg}]+|\Omega|^2},\nonumber \\
\Delta&=&\omega_p-\omega_{e_+f}(B=0).
\label{newsuscep1}
\end{eqnarray}
Here, $\Delta=2B$ is the detuning of the pump field from the transition 
$|3^2P_{1/2}; F=2,m_F=0\rangle \leftrightarrow |3^2S_{1/2}; F=2,m_F=0\rangle$
transition [see Fig.\ \ref{config1}], 
$\Omega= \vec{d}_{e_+f}.\vec{\cal E}_p/\hbar$ is the half of the pump Rabi 
frequency. The parameter $\Gamma_{fg}$ represents
the collisional dephasing between the states $|f\rangle$ and $|g\rangle$. In what
follows we use $\Gamma_{fg}=0$. The level $|f\rangle$ is Zeeman separated from the level $|3^2S_{1/2}; F=2,m_F=0\rangle$ by an amount of $3B$, whereas the levels $|e_\pm\rangle$ are separated by an amount $\mp B$.
These can be calculated from the Land\'e-g factor of the corresponding levels.
 The susceptibility $\bar{\chi}_-$ remains the same as
in (\ref{chi1}a). Note that in presence of  the control  field, the response of the
system is equivalent to a two-level system comprised of $(|e_-\rangle,|g\rangle)$ [for $\sigma_-$ component]
and a $\Lambda$-system [for $\sigma_+$ component] comprised of $(|e_+\rangle,|f\rangle,|g\rangle)$ 
connected via the common level $|g\rangle$.
{\narrowtext
\begin{figure}
\epsfxsize 5cm
\centerline{\begin{tabular}{cc}
\psfig{figure=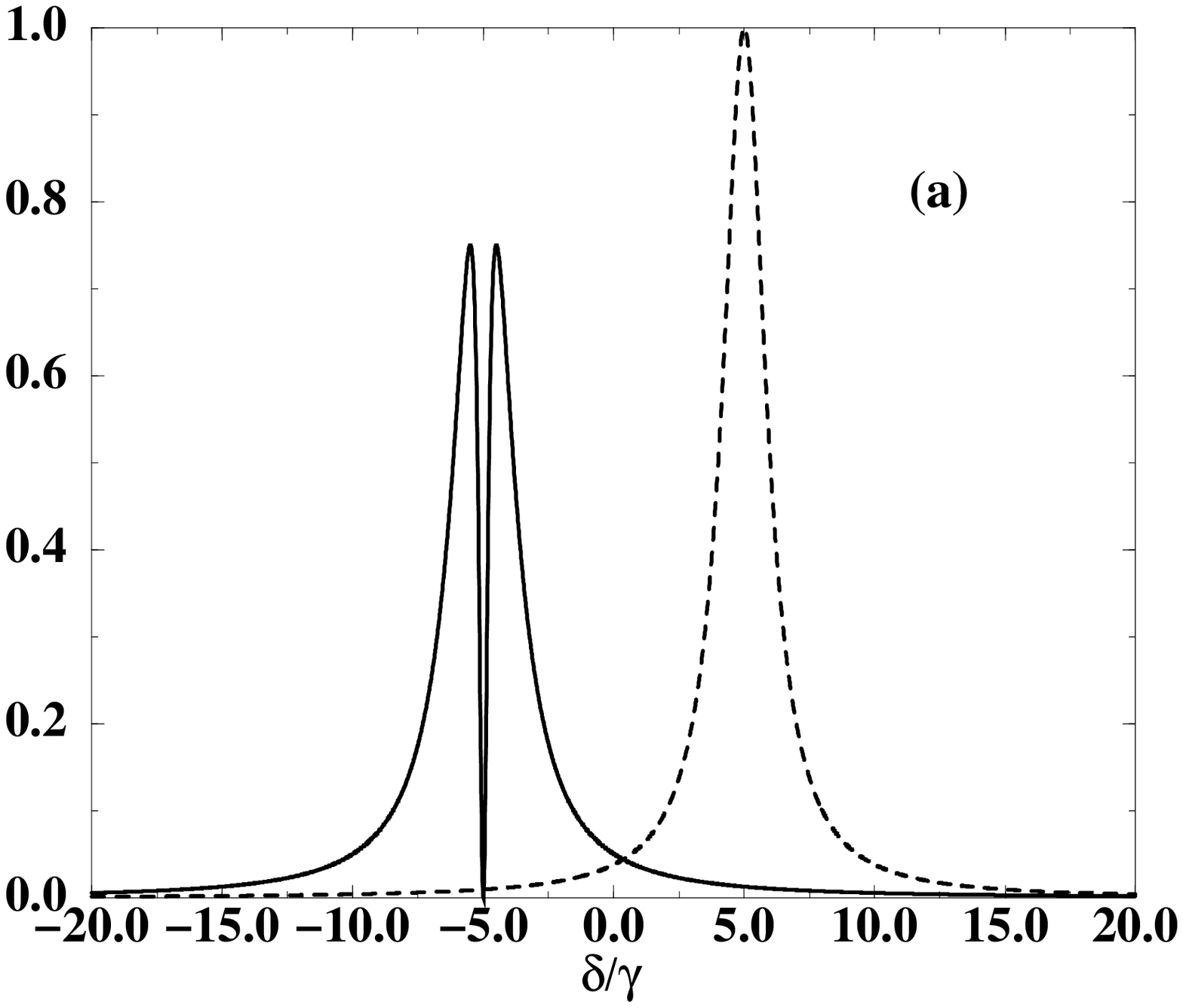,height=4cm}&
\psfig{figure=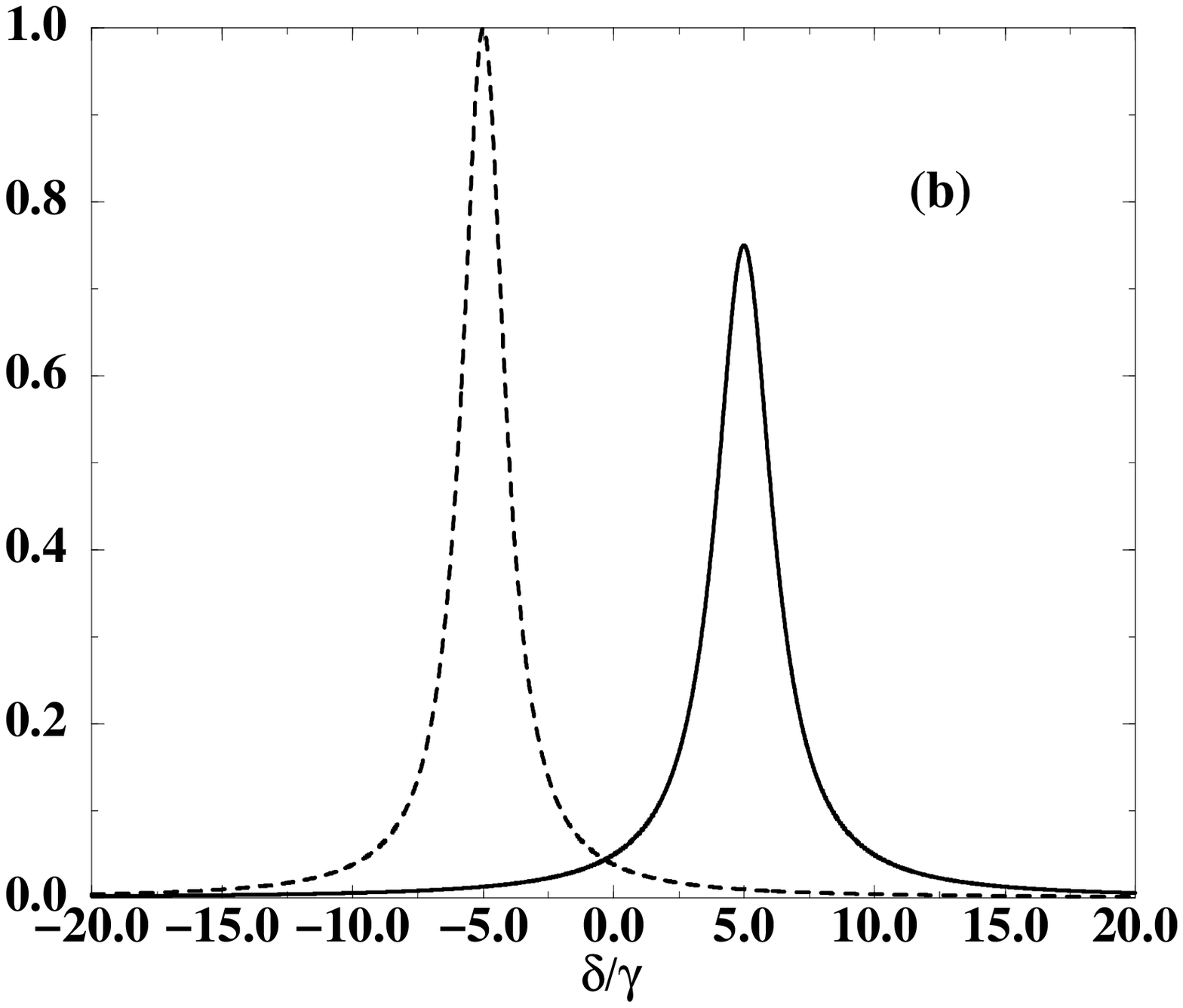,height=4cm}
\end{tabular}}
\caption{The variation of imaginary parts of the susceptibilities $\bar{\chi}_+$ (solid curve) and $\bar{\chi}_-$ (dashed curve) in units of $\alpha_0$ with probe
 detuning $\delta/\gamma$ are shown for the $\bar{\chi}_\pm (B)$ [(a)] and $\bar{\chi}_\pm (-B)$ [(b)].
The parameters used here are 
$\Omega=0.5\gamma$, $B=5\gamma$ corresponding to $105$ G, $\Gamma_{e_+g}=4\gamma/3$, $\Gamma_{fg}=0$, $\Gamma_{e_-g}=\gamma$, and $\Delta=2B$.}
\label{chiB1}
\end{figure}
}
It is clear that applying a coherent pump field, one can generate an EIT window 
at $\delta=-B$ (cf., $\Delta=2B$) for the $\sigma_+$ component 
[Im $(\bar{\chi}_+) = 0$]. On the
other hand, the absorption peak of the $\sigma_-$ component occurs at
$\delta=B$. Thus, this component suffers a little absorption 
[Im $(\bar{\chi}_-) \approx 0$] at $\delta=-B$ as the field is far detuned from the
 $|e_-\rangle\leftrightarrow |g\rangle$ transition as long as we choose the
magnetic field much larger than the width of the transition [see Fig.\ \ref{chiB1}(a)]. 
Thus the unpolarized probe field travels through the medium almost unattenuated.
 The transmittivity $T(B)=I_0(B)/I$ becomes almost unity at $\delta=-B$ as obvious 
from Eq.\ (\ref{outint1}) [see Fig.\ \ref{trans1}].
{\narrowtext
\begin{figure}
\epsfxsize 8cm
\centerline{
\epsfbox{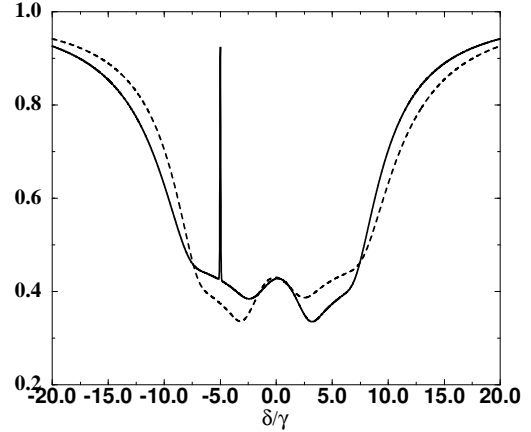}}
\caption{This figure shows the variation of the transmittivity $T(B)$ (solid 
curve) and $T(-B)$ (dashed curve) with respect to probe field detuning 
$\delta/\gamma$. The parameters used here are $N=10^{10}$ atoms cm$^{-3}$,
$\lambda=589$ nm, and $L=1$ cm. All the other parameters used are the same as in
Fig.\ \ref{chiB1}.}
\label{trans1}
\end{figure}
}
If now the direction of the magnetic field is reversed ($B\rightarrow -B$), 
then the corresponding susceptibilities for $\sigma_-$ and $\sigma_+$ polarizations become
\begin{mathletters}
\begin{eqnarray}
\bar{\chi}_-(-B)&=&\frac{-i\gamma\alpha_0}{i(\delta +B)-\Gamma_{e_-g}},\\
\bar{\chi}_+(-B)&=&\frac{-i\gamma\alpha_0[i(\delta-\Delta-3B)-\Gamma_{fg}]}{[i(\delta-B)-\Gamma_{e_+g}][i(\delta-\Delta-3B)-\Gamma_{fg}]+|\Omega|^2}\nonumber.\\
&&
\end{eqnarray}
\end{mathletters}
We continue to take the quantization axis as defined by the direction of the 
propagation of the electromagnetic field. Clearly now at $\delta=-B$, $\bar{\chi}_-(-B)$ has absorption peak and $\sigma_-$ component of the probe will be absorbed. If we continue to use $\Delta=2B$, i.e., if we keep the control laser frequencies fixed while we change the direction of the magnetic field, then $\bar{\chi}_+(-B)$ exhibits resonances at $\delta=3B\pm\sqrt{4B^2+\Omega^2}$, both of which are
far away from the point $\delta=-B$ unless we choose $\Omega^2=12B^2$. Clearly,
for $\delta=-B$ and $\Omega^2\neq 12B^2$, the $\sigma_+$ component of the probe will suffer very little absorption. This is in contrast to the behavior of the $\sigma_-$ component which will be attenuated by the medium. 
 Thus the output field would essentially have the contribution from the 
$\sigma_+$ component. The transmittivity $T(-B)=I_0(-B)/I$ of the medium 
decreases to about $1/2$. Thus by using EIT we can produce the result $T(B)\approx 2T(-B)$, i.e., we can alter the transmittivity of the medium 
by just reversing direction of the magnetic field. The equality (\ref{equal1})
no longer is valid and the medium behaves like a chiral medium. This becomes 
quite clear from the Fig.\ \ref{trans1}, at $\delta=-B$.

A quite a different result is obtained by choosing the parameter region differently. For the choice of the external field strength $\Omega= 2\sqrt{3}B$, and 
$\Delta=2B$, $\delta=-B$, the
$\sigma_+$ component gets absorbed significantly if the direction of the magnetic field is opposite to the direction of the propagation of the field. Thus $T(-B)$ becomes insignificant compared to $T(B)$ as shown in Fig.\ \ref{diffB1}.
For larger values of $B$, the result is shown in the inset. Here $T(B)/T(-B)$ is
in the range 2 to 3. Clearly, the case displayed in Fig.\ \ref{diffB1} is quite an unusual one. Such a large asymmetry in the dichroism of unpolarized light
is the result of the application of a coherent control field whose parameters 
are chosen suitably.

The behavior shown in the Fig.\ \ref{diffB1} is easily unterstood from the magnitudes of
the imaginary parts of the susceptibilties $\chi_\pm(\pm B)$. In the parameter domain 
under consideration, Im$[\chi_+(B)]=0$ (EIT); Im$[\chi_-(-B)=\alpha_0$,
 because the $\sigma_-$ component is on resonance for $\vec{B}$
antiparallel to the direction of propagation. Further as shown in the Fig.\ \ref{imchis},
in the region around $\Omega=2\sqrt{3}B$, Im$[\chi_-(B)] \ll \textrm{Im}[\chi_+(-B)]$. 
Thus both $\sigma_-$ and $\sigma_+$ components are absorbed if $\vec{B}$ is 
antiparallel $\vec{k}$, making the medium opaque as shown in the inset of Fig. 5.
The opacity disappears if direction of $\vec{B}$ is reversed (see Fig. 4, solid curve).
For values of $B$ away from the equality $\Omega=2\sqrt{3}B$, Im$[\chi_+(-B_)]$ 
decreases leading to an increase in the transmission $T(-B)$.
{\narrowtext
\begin{figure}
\epsfxsize 8cm
\centerline{
\epsfbox{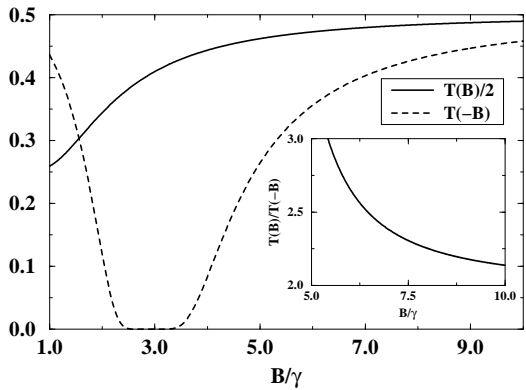}}
\caption{The transmittivities $T(B)$ and $T(-B)$  calculated at the value
$\delta=-B$ is plotted here with respect to $B/\gamma$, for $\Omega=10\gamma$ and 
$\Delta=2B$. The inset shows the magnetic field dependence
of the ratio $T(B)/T(-B)$ for the same parameters. All the other parameters are the same as in Fig.\ \ref{trans1}.}
\label{diffB1}
\end{figure}
}
{\narrowtext
\begin{figure}
\epsfxsize 8cm
\centerline{\epsfbox{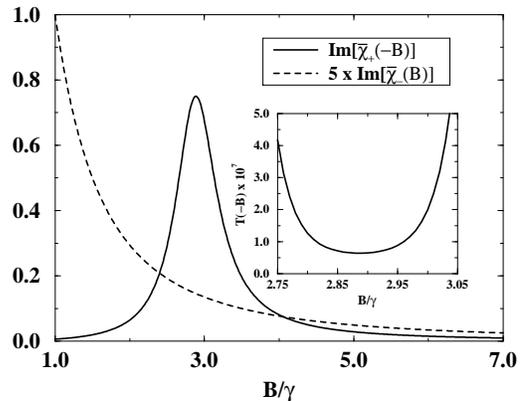}}
\caption{The variation of the imaginary parts of the susceptibilities $\chi_+(-B)$ and $\chi_-(B)$ in units of $\alpha_0$  with the magnetic field $B/\gamma$ 
for $\delta=-B$. The inset shows the variation $T(-B)$ with $B$ in the vicinity of
$\Omega=2\sqrt{3}B$. The parameters used here are the same as in Fig.\ \ref{diffB1}.}
\label{imchis}
\end{figure}}
\section{Large magnetic field reversal asymmetry in ladder system}
It may be recalled that there are many different situations where pump cannot be applied in  a
Lambda configuration. This, say, for example, is the case for $^{40}$Ca. The
relevant level configuration is
shown in Fig.\ \ref{config2} \cite{gaeta}. The level $|g\rangle$ $(|4s^2; j=0,m_j=0\rangle)$ is coupled to $|e_+\rangle$ $(|4s4p; j=1,m_j=+1\rangle)$ and
$|e_-\rangle$ $(|4s4p; j=1,m_j=-1\rangle)$ via the $\sigma_+$ and $\sigma_-$ components
 of the input unpolarized probe field, respectively. In this configuration,
the susceptibilities of the two circularly polarized components of the probe are given by
\begin{mathletters}
\begin{eqnarray}
\chi_+(B)&=&\frac{-i\gamma\alpha_0}{i(\delta-B)-\gamma},\\
\chi_-(B)&=&\frac{-i\gamma\alpha_0}{i(\delta+B)-\gamma},~\delta=\omega-\omega_{e_-g}(B=0),
\end{eqnarray}
\label{chi2}
\end{mathletters}
where, $\alpha_0=N|\vec{d}|^2/\hbar\gamma$, $N$ is the number density of the medium, $\delta$ is the detuning of the probe field with respect to the 
$|g\rangle \leftrightarrow |0\rangle$ $(|4s4p; j=1,m_j=0\rangle)$ transition, 
$|\vec{d}|$ is the magnitude of the dipole moment matrix element between the
levels $|e_+\rangle$ and $|g\rangle$, and $2\gamma$ is the decay rate from the 
levels $|e_+\rangle$ and $|e_-\rangle$ to the level $|g\rangle$.
Note that the imaginary parts of the susceptibilities (\ref{chi2}) are peaked 
at $\delta=B$ and $\delta=-B$, respectively, and clearly predict
perfect symmetry in the transmittivity of the medium upon reversal of the direction of the magnetic
field.
{\narrowtext
\begin{figure}
\epsfxsize 8cm
\centerline{
\epsfbox{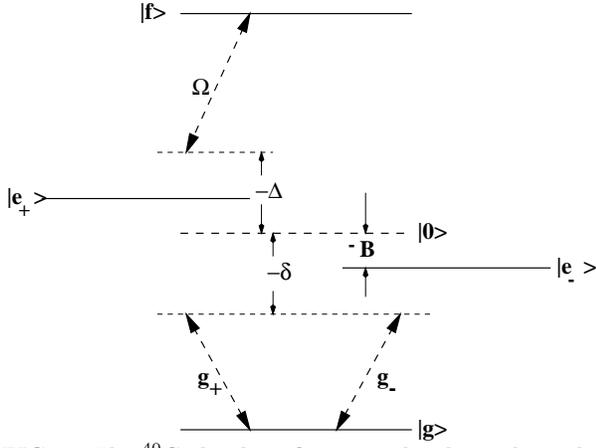}}
\caption{The $^{40}$Ca level configuration has been shown here. Here, $2g_\pm$ 
are the probe Rabi frequencies for the $\sigma_\pm$ components, $B$ is the 
magnetic field strength, $\Omega$ is the half of the pump Rabi frequency, 
$\delta$ and $\Delta$ are the respective detuning for the probe and the pump 
fields. These detunings are defined with respect to the energy separation 
between the levels $(|0\rangle,|g\rangle)$ and $(|0\rangle,|f\rangle)$, respectively.}
\label{config2}
\end{figure}
}
We will now show how one can use a coherent control field to create asymmetry
between $T(B)$ and $T(-B)$. We apply a coherent pump (\ref{pump1}) to couple $|e_+\rangle$ with
 a higher excited level $|f\rangle$ $(|4p^2; j=0,m_j=0\rangle)$ with Rabi frequency 
$2\Omega=2\vec{d}_{fe_+}.\vec{\cal E}_p/\hbar$.
The susceptibility for $\sigma_+$ component now changes to \cite{mor}
\begin{eqnarray}
\bar{\chi}_+(B)&=&\frac{-i\gamma\alpha_0[i(\Delta+\delta)-\Gamma]}{[i(\delta-B)-\gamma][i(\Delta+\delta)-\Gamma]+|\Omega|^2},\nonumber\\
\Delta&=&\omega_p-\omega_{fe_+}(B=0),
\label{newsuscep2}
\end{eqnarray}
where, $\Gamma=0.5(\lambda_{e_+g}/\lambda_{fe_+})^3\gamma=0.45\gamma$ is the spontaneous decay rate of the upper level $|f\rangle$ [cf., $\lambda_{e_+g}=422.7$ nm and $\lambda_{fe_+}=551.3$ nm], $\lambda_{\alpha\beta}$ is the wavelength of the transition between $|\alpha\rangle$ and $|\beta\rangle$,
$\Delta=-B$ is the detuning of the pump field from the $|f\rangle\leftrightarrow |0\rangle$ transition [see Fig.\ \ref{config2}]. 
Thus a transparency dip in the absorption profile of the $\sigma_+$ component 
at $\delta=B$ is generated and
the $\sigma_-$ component remains far-detuned from the corresponding transition,
 as shown in Fig.\ \ref{chiB2}(a). Note that the transparency for $\sigma_+$ is not
total which is in contrast to a Lambda system.
We display in the Fig.\ \ref{trans2} the behavior of the transmittivity
$T(B)$ of the medium as a function of the detuning. 
{\narrowtext
\begin{figure}
\epsfxsize 5cm
\centerline{\begin{tabular}{cc}
\psfig{figure=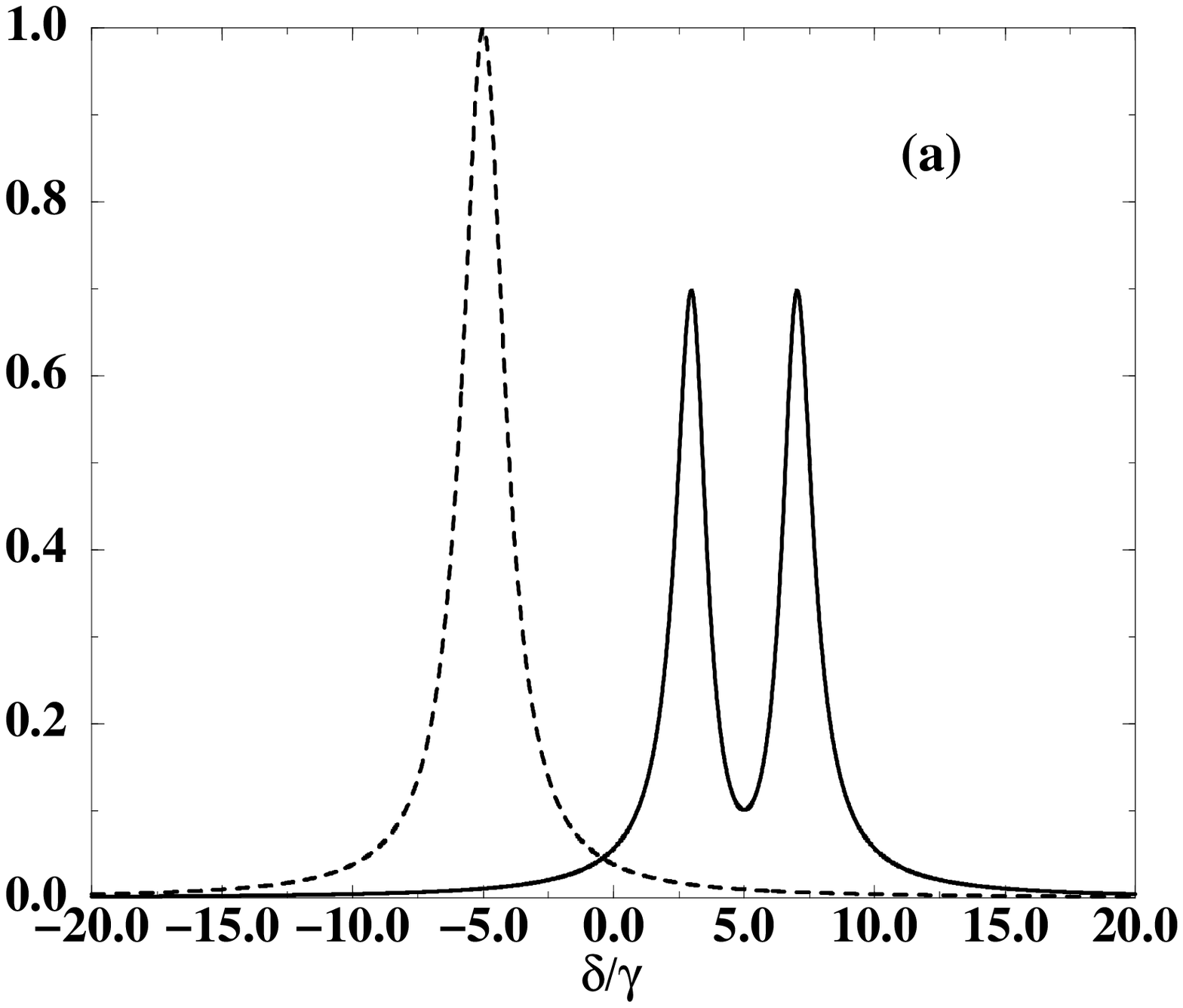,height=4cm}&
\psfig{figure=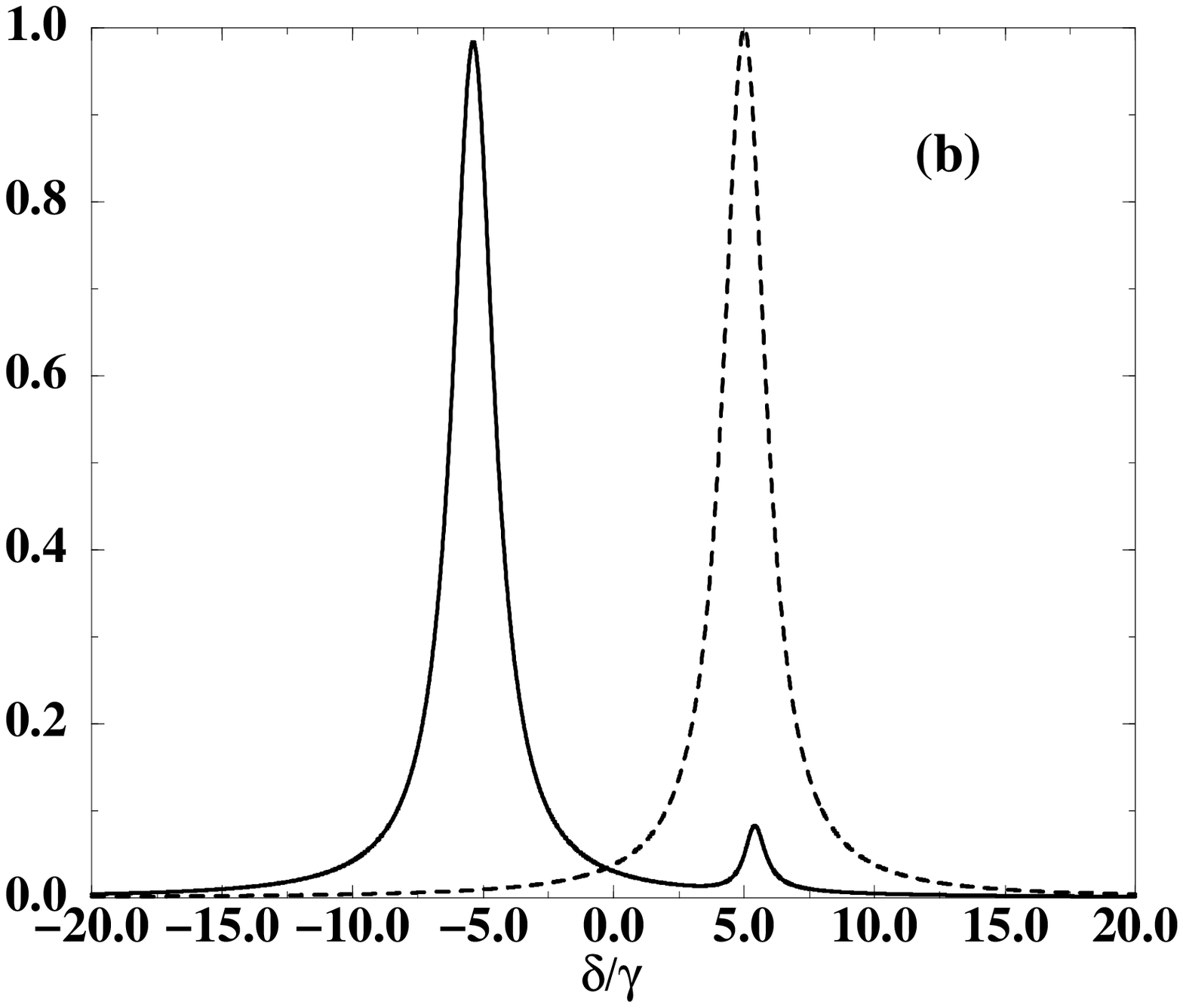,height=4cm}
\end{tabular}}
\caption{The variation of imaginary parts of the susceptibilities $\bar{\chi}_+$ (solid curve) and $\bar{\chi}_-$ (dashed curve) in units of $\alpha_0$ with respective to the probe detuning $\delta/\gamma$ are shown here for $\bar{\chi}_\pm(B)$ [(a)] and 
$\bar{\chi}_\pm(-B)$ [(b)]. The parameters used here are $\Omega=0.5\gamma$, $B=5\gamma$ corresponding to $123$ G, $\Gamma=0.45\gamma$, and $\Delta=-B$.}

\label{chiB2}
\end{figure}
}
Now upon reversal of the magnetic field direction, the $\sigma_+$ component gets
 detuned from corresponding transition and thereby suffers little absorption. 
The $\sigma_-$ component, being resonant with the corresponding transition, gets
largely attenuated inside the medium. This is clear from the 
Fig.\ \ref{chiB2}(b). Thus the contribution to the transmittivity $T(-B)$ 
comes primarily from the $\sigma_+$ component. The Fig.\ \ref{trans2} exhibits the
behavior of $T(-B)$ as the frequency of the probe is changed. In the region of EIT, $T(B)$ is several times of $T(-B)$.
{\narrowtext
\begin{figure}
\epsfxsize 8cm
\centerline{
\epsfbox{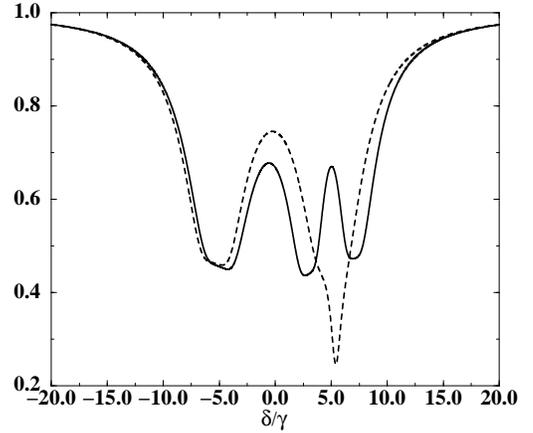}}
\caption{The variation for the transmittivities $T(B)$ (solid curve) and $T(-B)$(dashed curve) with the probe detuning $\delta/\gamma$ are shown in this figure. The parameters used here are $N=10^{10}$ atoms cm$^{-3}$, $\lambda_{e_+g}=422.7$ nm, $L=1$ cm, and the other parameters are the same as in Fig.\ \ref{chiB2}. }
\label{trans2}
\end{figure}
}

In Fig.\ \ref{diffB2}, we have shown how the ratio $T(B)/T(-B)$ calculated at 
$\delta=B$ is modified with change in the magnitude of the applied magnetic 
field for different control field Rabi frequencies. Note that for large $B$ and $\Omega$, this
ratio approaches the value of two, though for intermediate values it can exceed two.
{\narrowtext
\begin{figure}
\epsfxsize 8cm
\centerline{
\epsfbox{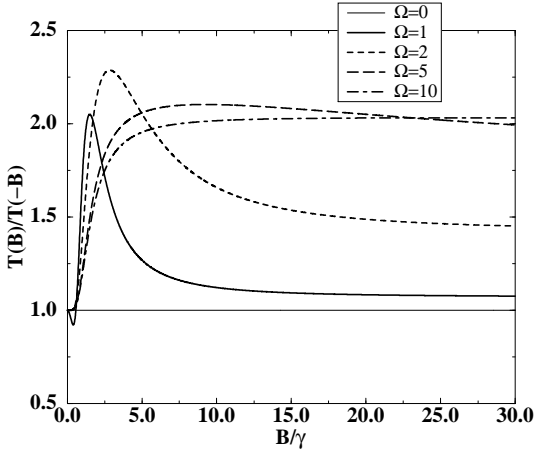}}
\caption{The variation of the ratio $T(B)/T(-B)$ calculated at $\delta=B$ with the magnetic field strengths $B/\gamma$ are shown in this figure for different values of $\Omega$. All the parameters are the same as in Fig.\ \ref{trans2}.}
\label{diffB2}
\end{figure}
}
\section{Magnetic field reversal asymmetry in the propagation of an unpolarized beam through a Doppler broadened medium}
We next consider the effects of Doppler broadening on the BRAS in a Lambda configuration.
We would like to find parameter regions where $T(B)$ and $T(-B)$ could differ 
significantly. We identify the spatial dependence of the pump ${\cal E}_p(z)=e^{ik_pz}$ and $\Delta_v=2B+k_pv_z$, where $v_z$ is the component of the atomic velocity 
in the direction of the propagation of the electric fields. We assume that 
the pump field propagates in the same direction $\vec{k}_p$ as the probe 
field wave vector $\vec{k}$ and we further take $k$ and $k_p$ to be approximately equal. 

We calculate the Doppler-averaged susceptibilities through the following relation:
\begin{equation}
\langle \bar{\chi}_\pm(v_z)\rangle = \int_{-\infty}^{\infty}\bar{\chi}_\pm(v_z)\sigma_D(v_z) dv_z,
\label{avg1}
\end{equation}
where, 
\begin{equation}
\sigma_D(v_z)=\frac{1}{\sqrt{2\pi \omega_D^2}}e^{-v_z^2/2\omega_D^2}
\label{Maxwell}
\end{equation}
is the Maxwell-Boltzmann velocity distribution at a temperature $T$ with the width $\omega_D=\sqrt{K_BT/M}$, $K_B$ is the Boltzmann constant, and $M$ is the 
mass of an atom.  

The integration (\ref{avg1}) results in a complex error function \cite{error}. However, to
have a physical understanding, we integrate (\ref{avg1}) by approximating
$\sigma_D$ by a Lorentzian $\sigma_L(v_z)$ of the width $\tilde{\omega}_D=2\omega_D\ln{2}$\cite{kash}:
\begin{equation}
\sigma_L(v_z)=\frac{\tilde{\omega}_D/\pi}{v_z^2+\tilde{\omega}_D^2}.
\label{lorentz}
\end{equation}
This leads to the following approximate results:
\begin{equation}
\langle\bar{\chi}_\pm(v_z)\rangle=\frac{\gamma\alpha_0}{k(v_\pm-i\tilde{\omega}_D)},
\label{approx}
\end{equation}
where,
\begin{eqnarray}
v_+&=&\frac{i\{[i(\delta+B)-\Gamma_{e_+g}]P+|\Omega|^2\}}{kP};\nonumber\\
P&=&i(\delta-\Delta+3B)-\Gamma_{fg},\\
v_-&=&\frac{i[i(\delta-B)-\Gamma_{e_-g}]}{k}.\nonumber
\end{eqnarray}
Here we have used the expressions (\ref{chi1}a) and (\ref{newsuscep1}) for $\bar{\chi}_\pm(v_z)$.
These susceptibilities (\ref{approx}) are used to calculate the transmittivities
at the point $\delta=-B$. In Fig.\ \ref{Dopp} we have shown the
corresponding variation of $T(B)$ and $T(-B)$ with $\delta/\gamma$. We find that the ratio $T(B)/T(-B)$ increases
to a value $\sim 1.6$, for a $6$ cm medium.
{\narrowtext
\begin{figure}
\epsfxsize 8cm
\centerline{
\epsfbox{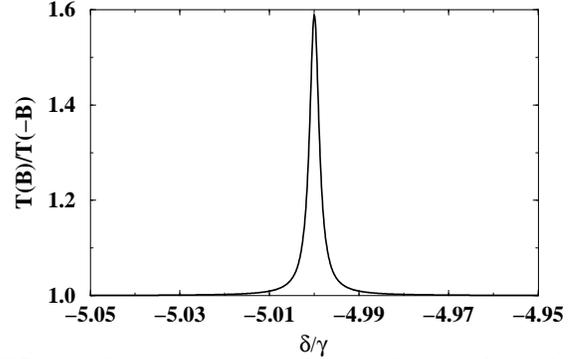}}
\caption{The variation in $T(B)/T(-B)$ with the probe
detuning $\delta/\gamma$ in a $6$ cm Doppler-broadened medium. All the other parameters are the same as in Fig.\ \ref{diffB1}. Note that $T(B)$ is the transmission
for $\vec{B}$ parallel to $\vec{k}$.} 
\label{Dopp}
\end{figure}
}
It is clear that, if we choose a longer medium in this case, the 
BRAS will be further enhanced. We have actually also carried out numerically the integration (\ref{avg1}). For the parameters of the Fig.\ \ref{Dopp}, the results do not change substantially.

\section{conclusions}
In conclusion, we have shown how one can make use of a coherent field to create large
BRAS in the propagation of an unpolarized light. We have discussed two different
system configurations. We have shown how EIT can be used very successfully to produce large BRAS. We have also analyzed the effect of Doppler broadening and found the interesting region of parameters with large asymmetry.

\end{multicols}
\end{document}